# Asymmetrical transport distribution function: skewness as a key to enhance thermoelectric performance

Jin-Cheng Zheng[(a,b)]*
(a) Department of Physics, Xiamen University, Xiamen 361005, China.
(b) Department of Physics and Department of New Energy Science and Engineering, Xiamen University Malaysia, Sepang 43900, Malaysia.
* Electronic mail: jczheng@xmu.edu.cn

**Abstract**

How to achieve high thermoelectric figure of merit is still a scientific challenge. By solving the Boltzmann transport equation, thermoelectric properties can be written as integrals of a single function, the transport distribution function (TDF). In this work, the shape effects of transport distribution function in various typical functional forms on thermoelectric properties of materials are systematically investigated. It is found that the asymmetry of TDF, characterized by skewness, can be used to describe universally the trend of thermoelectric properties. By defining symmetric and asymmetric TDF functions, a novel skewness is then constructed for thermoelectric applications. It is demonstrated, by comparison with ab initio calculations and experiments, that the proposed thermoelectric skewness not only perfectly captures the main feature of conventional skewness, but also is able to predict the thermoelectric power accurately. This comparison confirms the unique feature of our proposed thermoelectric skewness, as well as its special role of connection between the statistics of TDF and thermoelectric properties of materials. It is also found that the thermoelectric performance can be enhanced by increasing the asymmetry of TDF. Finally, it is also interesting to find that the thermoelectric transport properties based on typical quantum statistics (Fermi-Dirac distributions), can be well described by typical shape parameter (skewness) for classical statistics.

## 1. Introduction

Recently, the thermoelectric energy conversion technology attracts more and more attentions due to its ability to convert heat into electricity or vice versa, with extreme reliability and without producing greenhouse gas emissions [1-6]. Therefore, thermoelectric devices can be used for power generation or refrigeration devices [1-8]. The energy conversion efficiency of thermoelectric devices is evaluated in terms of a dimensionless thermoelectric figure of merit, *ZT*, which depends on the transport

*Note: Equation (20) in this version incorporates the correction published in Research 2026, Article 1317 (DOI: 10.34133/research.1317).*

coefficients of the constituent materials, and can be defined by: [6]

$$ZT = \frac{\sigma S^2 T}{\kappa_e + \kappa_l} \tag{1}$$

where σ is the electrical conductivity, $S$ is the Seebeck coefficient (thermoelectric power), $T$ is the mean operating temperature, $\kappa_e$ and $\kappa_l$ are electronic and lattice thermal conductivities, respectively. The term $\sigma S^2$, the so-called power factor, characterizes the pure electronic contribution to $ZT$. A general strategy for improving thermoelectric efficiency is to increase the electrical conductivity and Seebeck coefficient, (equivalently, to increase the power factor), but to decrease the thermal conductivity [5]. However, in real materials, the improvement of thermoelectric efficiency is not straightforward but rather difficult, due to the strong coupling among three associated components, $\sigma$, $S$, and $\kappa$. For example, it is well known that, compared with insulator, metals have very good electrical conductivity and one may expect from Eq. (1) that metals can be good thermoelectrics. In fact, most metals are very poor thermoelectric materials due to their very small Seebeck coefficient and large thermal conductivity. For insulators with large band gaps, they have very large Seebeck coefficients. But their ultra small electrical conductivity significantly reduces the power factor and thus results in a very small $ZT$ [5]. The optimal thermoelectric materials are therefore located in the region near the crossover between semiconductor and metal, and the mechanism for enhancement of thermoelectric performance is thus desirable to be further studied.

Different approaches have been reported to increase the thermoelectric figure of merit. To reduce the lattice thermal conductivity, increased phonon scattering by defects and nanocrystals [9, 10, 11], by layer structured $PbSnS_2$ or doping [12] in PbTe based thermoelectric materials [13], by the hierarchical phonon scattering centers [14, 15] and interfaces [16, 17, 18], as well as the nanowires [19], has been reported. The electronic conductivity, Seebeck coefficient and power factors have been studied by investigating the effects of pressure [20, 21], strains [22, 23, 24], defects [25], doping [12, 26], polytypes [27], superlattices [28, 29], and reduced dimension or nanostructures [30, 31]. The effects of symmetry of crystal structure on thermoelectric properties [32,33] and the effects of Aubry–Andre–Harper modulation on nano scale thermoelectrics [34] have been investigated. Advancements have been made recently for the $Bi_2Te_3$ nanofilms [28, 35], other chalcogenides based materials [26, 36, 37, 38], $CaMg_2Bi_2$-based compounds [39] or $Mg_3Sb_{2-x}Bi_x$ alloys [40, 41], Half-Heusler materials [42], semimetals [43], and organic materials [44, 45, 46, 47]. An interesting increasing trend can be observed by utilizing the machine learning method to assist research in energy materials [48, 49, 50] including thermoelectric materials [42,51].

All the aforementioned approaches are aiming to optimize some particular components of thermoelectric properties. Unfortunately, these components are coupled with each other, and are not easy to be optimized simultaneously. A question then arises: is there a unified picture to describe thermoelectric properties? We note that $\sigma$, $S$, and $\kappa_e$ can be expressed in terms of integrals of a single function, the transport distribution function (TDF), which consists of carrier velocity, relaxation time and density of states (DOS). Therefore, the essential step to search for good thermoelectrics is to find the optimal TDF, which gives high thermoelectric figure of merit. Although the transport distribution function in the delta function form has been shown to give the best thermoelectric performance mathematically [3], the shape effect of TDF in other mathematical form on thermoelectric properties has not been well studied. Moreover, in real materials, the TDF is usually having complex and broadened shape, rather than being an ideal delta function with a sharp peak. The universality of the shape effect of the TDF on thermoelectric performance becomes more and more important and emergent for the search of or the optimal design for efficient thermoelectric materials.

In this work, a systematical investigation of the shape effects of TDF in a varity of typical functional forms on thermoelectric properties of materials is carried out, in order to provide a unified picture with emphasis on asymmetrical TDF, characterized by skewness, which describes the universal features of thermoelectric power, power factor and *ZT*, so as to enrich our understanding of electronic origin for improving thermoelectric efficiency. The basic theory of thermoelectricity based on the Boltzmann transport equation is briefly introduced first, and the effects of TDF in terms of location parameter and scale parameter on thermoelectric performance are then presented. After that, detailed discussions are provided focusing on the asymmetric feature of TDF, characterized by skewness, as an important key to describe the thermoelectric transport properties. We have proposed a novel skewness, named as thermoelectric skewness, constructed by the symmetric and asymmetric functions of TDF, and compared our thermoelectric skewness with conventional skewness (third moment), then presented its applications combined with ab initio results and experiments.

## 2. Theory of thermoelectricity based on the Boltzmann transport equation

By solving the Boltzmann transport equation (BTE), the three properties $\sigma$, $S$, and $\kappa_e$ mentioned above can be written as integrals of a single function, the TDF[3,52]:

$$\sigma(T,\mu) = \int_{-\infty}^{+\infty} \left(-\frac{\partial f_\mu(T,\varepsilon)}{\partial \varepsilon}\right) \sigma(\varepsilon) d\varepsilon \tag{2}$$

$$S(T,\mu) = -\frac{1}{eT\sigma(T,\mu)} \int_{-\infty}^{+\infty} \left(-\frac{\partial f_\mu(T,\varepsilon)}{\partial \varepsilon}\right) \sigma(\varepsilon)(\varepsilon-\mu) d\varepsilon \tag{3}$$

$$\kappa_e(T,\mu) = \frac{1}{e^2 T} \int_{-\infty}^{+\infty} \left(-\frac{\partial f_\mu(T,\varepsilon)}{\partial \varepsilon}\right) \sigma(\varepsilon)(\varepsilon-\mu)^2 d\varepsilon - T\sigma(T,\mu) S^2(T,\mu) \tag{4}$$

where $\sigma(\varepsilon)$ is the TDF, $f_\mu(T,\varepsilon)$ is the Fermi-Dirac distribution function, $\mu$ is the chemical potential, and $T$ is the temperature. The derivative of $f_\mu(T,\varepsilon)$ is expressed as

$$-\frac{\partial f_\mu(T,\varepsilon)}{\partial \varepsilon} = \frac{exp\left(\frac{\varepsilon-\mu}{k_B T}\right)}{k_B T\left[exp\left(\frac{\varepsilon-\mu}{k_B T}\right)+1\right]^2} \tag{5}$$

By defining the dimensionless variable $x = (\varepsilon\text{-}\mu)/(k_B T)$, constants $\sigma_0 = e^2/(\hbar a_0) \sim$ 46, 000 $(\Omega\ \text{cm})^{-1}$ and $S_0 = k_B/e \sim 87\ \mu\text{V K}^{-1}$, putting other material dependent parameters into one single variable, $\lambda = \kappa_l/[T\sigma_0(k_B/e)^2] = \kappa_l/[T\sigma_0(S_0)^2]$, and using a dimensionless TDF:

$$g(x) = \hbar a_0 \sigma(\mu + x k_B T) \quad (6)$$

one can express the transport coefficients in terms of dimensionless variable and dimensionless integrals.

The reduced electrical conductivity $\sigma_r$ can be written as,

$$\sigma_r = \sigma/\sigma_0 = J_0, \quad (7)$$

the reduced thermoelectric power (Seebeck coefficient) $S_r$ becomes,

$$S_r = S/S_0 = J_1/J_0, \quad (8)$$

the reduced power factor $P_r$ is,

$$P_r = (\sigma S^2)/(\sigma_0 S_0^2) = J_1^2/J_0, \quad (9)$$

and the reduced electronic thermal conductivity $\kappa_{e,r}$ can be expressed as,

$$\kappa_{e,r} = \kappa_e/(T\sigma_0 S_0^2) = (J_2 - J_1^2/J_0), \quad (10)$$

where the dimensionless integrals $J_n$ can be written as:

$$J_n = \int_{-\infty}^{+\infty} D_n(x) g(x) dx \quad (11)$$

and the distribution density function (DDF), $D_n(x)$, is defined by

$$D_n(x) = \frac{exp(x)}{[exp(x)+1]^2} x^n \quad (12)$$

The shape of $D_n(x)$ is illustrated in Fig. 1. It is clearly seen that $D_0(x)$ has a maximum at $x = 0$, and then decreases when $x$ is larger or smaller than 0. It drops quickly to near zero when $x \sim \pm 5$, i.e., $(\varepsilon\text{-}\mu) \sim \pm 5k_B T$.

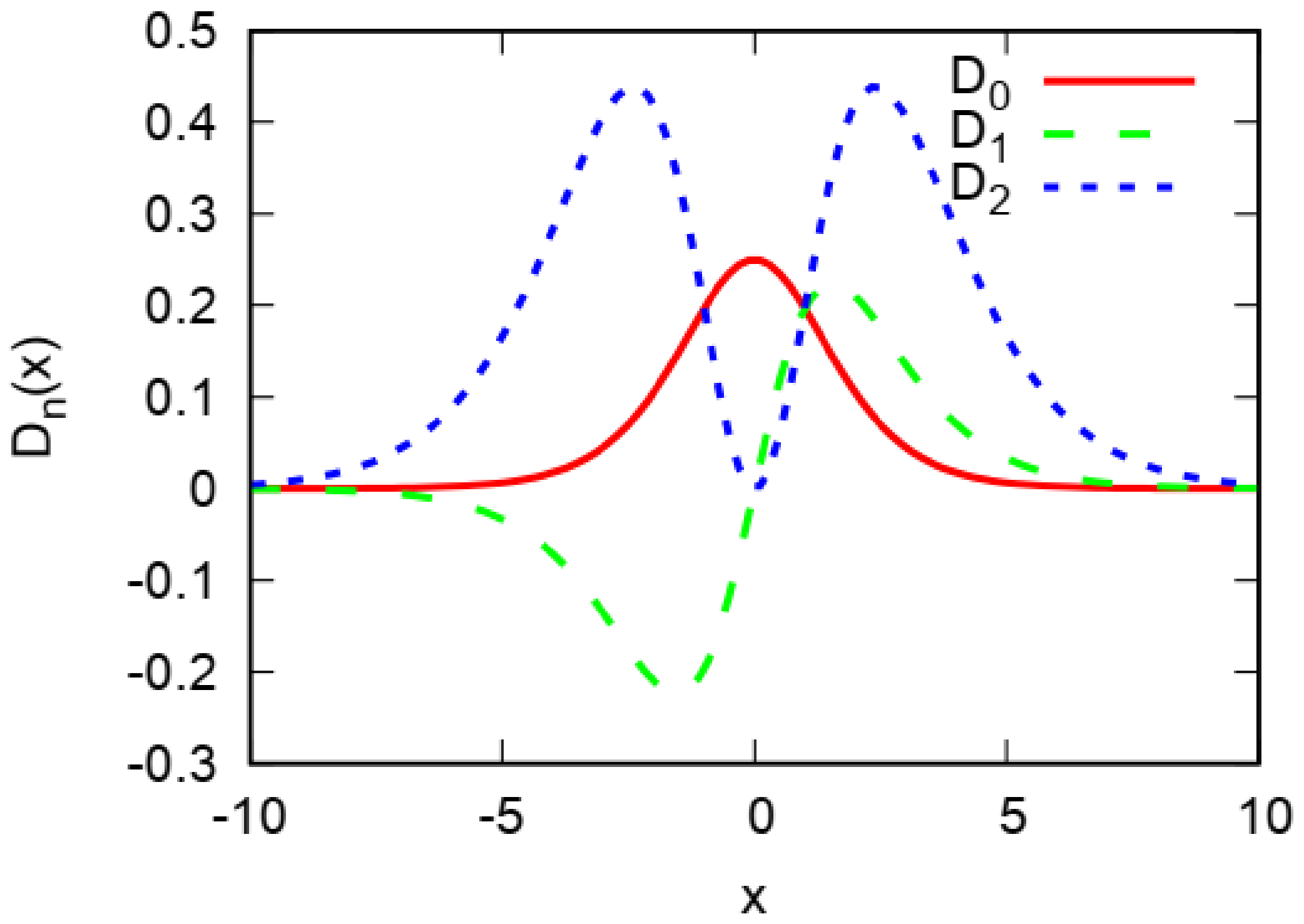

Fig. 1. The distribution density function $D_n(x)$ as a function of $x$ for different $n$.

Finally, the figure of merit can be obtained subsequently,

$$ZT = \frac{\sigma_r S_r^2}{\kappa_{e,r}+\lambda} = \frac{J_1^2/J_0}{J_2 - J_1^2/J_0+\lambda} \tag{13}$$

If we further express $\lambda$ in terms of the reduced temperature $T_r=T/T_0$ with Room temperature $T_0 = 300$ K, the reduced lattice thermal conductivity $\kappa_{l,r} = \kappa_l/\kappa_0$ with $\kappa_0 = 1$ $Wm^{-1}K^{-1}$, and the dimensionless constant $\beta_0=\kappa_0/[T_0\sigma_0(S_0)^2]= 0.0976$, one can see that $\lambda$ is actually a dimensionless quantity, too:

$$\lambda = \beta_0 \kappa_{l,r}/T_r. \tag{14}$$

For $T$ = Room temperature (RT), $T_r = 1$, we have $\lambda = \beta_0\kappa_{l,r} = 0.0976\kappa_{l,r}$. The expression of $ZT$ in Eq. (13) is different from the conventional expression like Eq. (1). The advantage of expressing $ZT$ in form of Eq. (13) is that every component in the equation remains dimensionless through using reduced quantity. Thus the formula of $ZT$ is in a neat and simple format. In this way, it is obvious that $ZT$ is only determined by the material dependent parameter $\lambda$ (phonon part) and the dimensionless TDF $g(x)$ (electron part). Because the reduced lattice thermal conductivity $\kappa_{l,r}$ (hence $\lambda$) is mainly determined by lattice contribution, the electronic mechanism for improving $ZT$ is to optimize TDF $g(x)$. The shape of $g(x)$ is therefore an important key factor for optimizing thermoelectric properties. In the following, we will examine the shape effects of TDF $g(x)$ in several typical distribution forms on thermoelectric performance. We will try to link the form of TDF as closely with real materials as possible in order to enhance the applicability of the theory.

## 3. The effects of location and scale parameters of TDF on thermoelectric performance

The *location parameter* of a probability distribution in statistics is a scalar- or vector-valued parameter $x_0$, which determines the shift of the distribution. For TDF, the location parameter is associated with chemical potential $\mu$, because the energy in TDF is relative to chemical potential, ($\varepsilon$-$\mu$). In the dimensionless expression of TDF, the location parameter is expressed in the variance of $x$. The *scale parameter* is a special kind of numerical parameter of a parametric family of probability distributions, in probability theory and statistics. The larger the scale parameter, the more spread out the distribution. The inverse scale parameter, or the so-called rate parameter, which is the reciprocal of the scale parameter, is often used in some families of distributions. The location parameter and scale parameter are also considered together, forming a *location–scale family*, which is a family of probability distributions parametrized by a location parameter and a non-negative scale parameter. The thermoelectric performance is therefore associated with the location parameter (peak positions or modes) and scale parameter of TDF.

The simplest extremely case is that the TDF is only depending on *scale parameter*, a constant $g_0$, namely, $g(x) = g_0$, but there is no location parameter. For a constant TDF with only scale parameter, the electrical conductivity is proportional to the scale parameter, *i.e.*, the constant $g_0$, but the thermoelectric components such as $S_r$ and $P_r$ are all zero, and thus no thermoelectric effect can be observed (see Supplemental materials (section S1) for more details).

A typical but extreme example for examining effects of *location-scale family* of TDF on thermoelectric properties is a single peak with Dirac delta function form. For the normalized case of TDF in the function form of Dirac delta function, the thermoelectric transport functions are determined by two constants, namely, the *location parameter* $x_0$ and the *scale parameter*, as had been shown in the earlier work of Mahan and Sofo [3], as well as in the recent work of Fan, Wang and Zheng [53]. The TDF in the single Dirac delta function form can be expressed as $g(x) = g_0\delta(x - x_0)$, where $g_0$ is constant. This TDF is actually a Delta shape peak located at the $x = x_0$ position away from the Fermi energy level. The reduced electronic thermal conductivity is 0, independent on location-scale family (see Supplemental materials (section S2) for more details). The reduced thermoelectric power is solely depending on the location parameter, $x_0$, but it is independent on the scale parameter, $g_0$. While the rest thermoelectric components, such as the reduced electric conductivity, the reduced power factor and the *ZT* are up to the scale parameter $g_0$, and the location parameter $x_0$, as well as the function of location parameter, $D_0(x_0)$.

For relating the Delta-function-like TDF to real materials, the rare-earth metal compounds are often referred to due to their very narrow *f* orbitals [3]. However, here, it is important to point out that cautions should be taken when linking to rare-earth metal compounds. As mentioned at the beginning of this article, the Boltzmann transport equation is used to describe the thermoelectric transport, which indicates that the systems with diffusive transport are considered. While for rare-earth metal compounds, if the narrow *f* orbitals are involved in thermoelectric transport, the localized orbital with possible strongly correlated features should be taken into account. In other words, to accurately describe electron strongly correlated system, the Kubo theory [54] should be used.

In real materials with itinerant electrons, the TDF usually is not as localized as Delta function; it may have finite height with finite spread. Therefore, the TDF with typical general location-scale family can be described with Gaussian function form,

$$g(x) = \frac{1}{b\sqrt{2\pi}} exp\left[-\frac{(x-x_0)^2}{2b^2}\right]. \tag{15}$$

The location parameter $x_0$ defines the peak position of the Gaussian function, and the scale parameter $b$ characterizes the width of the peak. Fan, Wang and Zheng [53] had discussed how to optimize the *ZT* with TDF being in the Gaussian function form with normalized case or bounded condition.

Based on the investigation of normalized Gaussian TDF, the good thermoelectrics should have the TDF with the characteristics of two factors (see Supplemental materials (section S3) for more details). One is the optimal location parameter, which should be about $x_0 = 2.4$. This equally means that the peak of TDF related to Fermi energy or chemical potential should be about $2.4k_BT$ (~ 60 meV at room temperature). The other factor is the scale parameter, for which, the smaller $b$ the better thermoelectric performance. This is indeed the optimization strategy for achieving high thermoelectric performance with the condition of assuming TDF is normalized. The normalized Gaussian TDF can provide some guidance for searching better thermoelectrics.

However, it is not fully valid because in the real materials, the DOS is not normalized, rather, the DOS is bounded by the number of electrons in the orbitals. With the bounded condition, we have shown previously that the optimal TDF is a rectangular-shape distribution [53]. For both normalized and bounded cases, the location parameter $x_0$ is still valid, but the scale parameter is depending on the normalizing condition. Obviously, we need to find other factor that can characterize the thermoelectric performance in a universal manner, and cover all the aforementioned aspects.

Although we try to make Gaussian TDF as general as possible, there is still something missing. The location parameter and scale parameter are describing the center and spread of a distribution. The characterization of the shape (or form) of distribution, which requires higher order of moments, is missing.

We note that, in real materials, the TDF above Fermi level (or chemical potential) can be quite different from the TDF below the Fermi level. This is actually quite popular case for materials ranging from metal to semiconductor. In fact, the optimal TDFs such as normalized Dirac delta function [3,53] or bounded rectangle TDF [53], are asymmetric about the chemical potential. Therefore, we propose in this work that the asymmetry of the TDF may be more general, and possible to be the universal characterization for thermoelectric properties. The asymmetry of the TDF is associated with the shape (or form) of distribution.

## 4. Skewness as a key to enhance thermoelectric performance

The shape of a probability distribution can be described by so-called shape parameter or form parameter, which is a kind of numerical parameter of a parametric family of probability distributions [55]. It is often measured quantitatively by the method of moments. The skewness ($3^{rd}$ moment) or kurtosis ($4^{th}$ moment) is often used to describe the shape. Especially, the skewness is a measure of the asymmetry of the probability distribution of a real-valued random variable about its mean.

The typical function forms of TDF, including constant TDF[3], delta function TDF[3,51], Gaussian TDF[53] and rectangle TDF[53] are all symmetric and having zero skewness about their mean. To discuss TDF in a more general manner, we need to choose typical distribution function with adjustable higher order moments, especially the skewness, to describe the effects of asymmetry of the TDF on the thermoelectric properties. We found that the Beta distribution function is well suited for this purpose.

The TDF using Beta distribution can be expressed as,

$$g(x,\alpha,\beta) = \frac{x^{\alpha-1}(1-x)^{\beta-1}}{B(\alpha,\beta)}, \tag{16}$$

where $B(\alpha,\beta) = \frac{\Gamma(\alpha)\Gamma(\beta)}{\Gamma(\alpha+\beta)}$, which is a normalization constant to ensure that the total probability is 1. Here, $\Gamma(z)$ is the Gamma function. The Beta distribution is more general than other aforementioned TDFs, and it may take a wide variety of different shapes depending on the values of the two parameters $\alpha$ and $\beta$ within the range of [0,1], as shown in Fig. 2. It can have symmetric shapes when $\alpha = \beta$ (see Fig. 2(a)), taking U-shaped distribution ($\alpha = \beta < 1$), 2-point Bernoulli distribution with equal probability 1/2 at each Dirac delta function end $x$=0 and $x$=1 ($\alpha = \beta \rightarrow 0$), the arcsine distribution ($\alpha = \beta$=1/2), uniform distribution ($\alpha = \beta$=1), the semi-elliptic distribution ($\alpha = \beta$=3/2), the parabolic distribution ($\alpha = \beta$=2), the bell-shaped distribution ($\alpha = \beta$>2), and the 1-point Degenerate distribution with a Dirac delta function spike at the midpoint $x$=0.5 with probability 1 and zero probability everywhere else ($\alpha = \beta \rightarrow \infty$). For the cases of $\alpha \neq \beta$ (see Fig. 2(b)), the beta density function is skewed. It can take U-shaped distribution or J shaped distribution with positive skew for $\alpha < \beta$, and negative skew for $\alpha > \beta$. The beta distribution has wide applications in statistics, physics, project management, and so on. We have successfully applied a beta-like distribution,

$$P_l(x) = \binom{4}{l} x^l (1-x)^{4-l} \tag{17}$$

as the statistic weight to account for the possibility of the $l$ short region ordered structure taking place in the semiconductor alloy $A_xB_{(1-x)}$. The band energy and the band offset as well as the formation energy for disorder alloys have been calculated using cluster expansion with above beta-like distribution [56-60].

The thermoelectric properties are thus can be obtained using the integrations listed in equations (7-13), in terms of parameters $\alpha$ and $\beta$. The skewness of the beta distribution can be expressed with third standardized moment,

$$\gamma_3 = E\left[\left(\frac{X-\mu_m}{\mathrm{SD}}\right)^3\right] = \frac{2(\beta-\alpha)\sqrt{\alpha+\beta+1}}{(\alpha+\beta+2)\sqrt{\alpha\beta}}. \tag{18}$$

Where $\mu_m$ is the mean, SD is the standard deviation, and E is the expectation operator. The skewness can be positive or negative.

As we discussed before, the thermoelectric TDF has limited effective energy range about $x \sim \pm 5$ , namely, $(\varepsilon-\mu) \sim \pm 5k_BT$ (see for example, Fig. 1), the finite range [0,1] of beta function can be scaled and shifted to be within the effective energy range for thermoelectric TDF. For convenient illustration, we have scaled beta TDF by 2.4, so that the case of delta function located at $x$=0 or $x$=1 can be reproduced. The chemical potential $\mu/(k_BT)$ is set to be the middle of interval, 0.5.

Firstly, we investigate the thermoelectric properties for symmetric TDF. For the symmetric Beta distribution, the parameters $\alpha = \beta$, thus the mean = 1/2, and the skewness equals to zero. The calculated thermoelectric power (Seebeck coefficient), power factor, and $ZT$ are found to be all zero, closely related with skewness. This is very interesting and has been shown for the first time that, ***for any form of TDF, if it is symmetric about its mean, the thermoelectric power (Seebeck coefficient), power factor, and ZT are zero for the case of the chemical potential μ/( k*<sub></sub>**B**T) being set to the mean of the distribution.*** This important conclusion can be verified through the first order of distribution density function (DDF), $D_1(x)$ (as shown in Fig. 1), and then the related dimensionless integrals $J_1$. Because the $D_1(x)$ is perfectly asymmetric about its mean, and just opposite for positive and negative $x$, namely, $D_1(x) = -D_1(-x)$, any integration (such as $J_1$) in the interval $[-x : x]$ will result in zero value. Any thermoelectric quantities that are proportional to $J_1$ will go to zero too.

Then we consider the cases of TDF with asymmetric shapes for $\alpha \neq \beta$. We start from the asymmetric linear Beta TDF with slope=2, namely, $g(x,\alpha=2, \beta=1) = 2x$ that is away from uniform distribution $g(x,\alpha=1, \beta=1) = 1$. The reduced electrical conductivity $\sigma_r(2,1)$ is about 0.224 (the chemical potential $\mu/(k_B T)$=0.5), which is almost the same as $\sigma_r(1,1)$, but the reduced thermoelectric power (Seebeck coefficient) increases from zero to $S_r(2,1)$ ~0.082, the power factor and *ZT* are enhanced from 0 to 0.030, and 0.179, respectively. Similarly, a significant magnitude of skewness is found to increase from 0 to 0.566, indicating some relationship between the thermoelectric power and the skewness of the TDF. A negative skew -0.566 is found for $g(x,\alpha=2, \beta=1)$ and positive skew is 0.566 for $g(x,\alpha=1, \beta=2)$. The thermoelectric power (Seebeck coefficient) have the same sign to skewness.

We further explore the relationship between skewness and thermoelectric power and power factor as well as *ZT* for the TDF similar to parabolic distribution ($g(x,\alpha=2,\beta=2)$) but with some degree of skewness ($g(x,\alpha=2,\beta)$). Fig. 3 plots the thermoelectric properties together with the skewness of TDF in the form of ($g(x,\alpha=2,\beta\geq2)$). The absolute value of Seebeck coefficient increases monotonically with parameter $\beta$. It can be clearly seen that the skewness of TDF is also increasing with parameter $\beta$ monotonically, very similar with the trend of Seebeck coefficient. The power factor and *ZT* are all positive due to the square of the thermoelectric power term in the formula, and therefore, we compare with the same order of the skew, that is, $\gamma_3^2$. As shown in Fig. 3(c), the power factor can be fitted closely with square of skew with a factor of 0.151, namely, PF ~ $0.151\gamma_3^2$. The variance of square of skew can also well capture the trend of *ZT* as a function of $\beta$ (Fig. 3(d)). The close relationship between ZT and square of skew ($\gamma_3^2$) for Beta TDF with more general cases (for any value of $\alpha$ and $\beta$) can be observed in Fig. 4. This result is striking and surprising in two folds. Firstly, this work presented for the first time that ***the asymmetry of TDF, characterized by skewness, is a universal parameter to describe the trend of thermoelectric properties for the shape of TDF***. Secondly, we have shown for the first time that ***the thermoelectric transport properties including thermoelectric figure of merit, which are based on typical quantum statistics (Fermi-Dirac distributions), can be well described by typical shape parameter (skewness) for classical statistics***.

Our results indicate that it is possible to choose suitable parameters from classical statistics to capture some important features of quantum statistics, thus the quantum statistics and classical statistics can be linked or bridged by some typical parameters that characterizing the shape of distribution. This may suggest that large amount of information or tools in classical statistics or probability theory may be possible to be applied for quantum statistics and used for optimization of important properties (such as the thermoelectric power, the power factor, and *ZT*) of quantum systems.

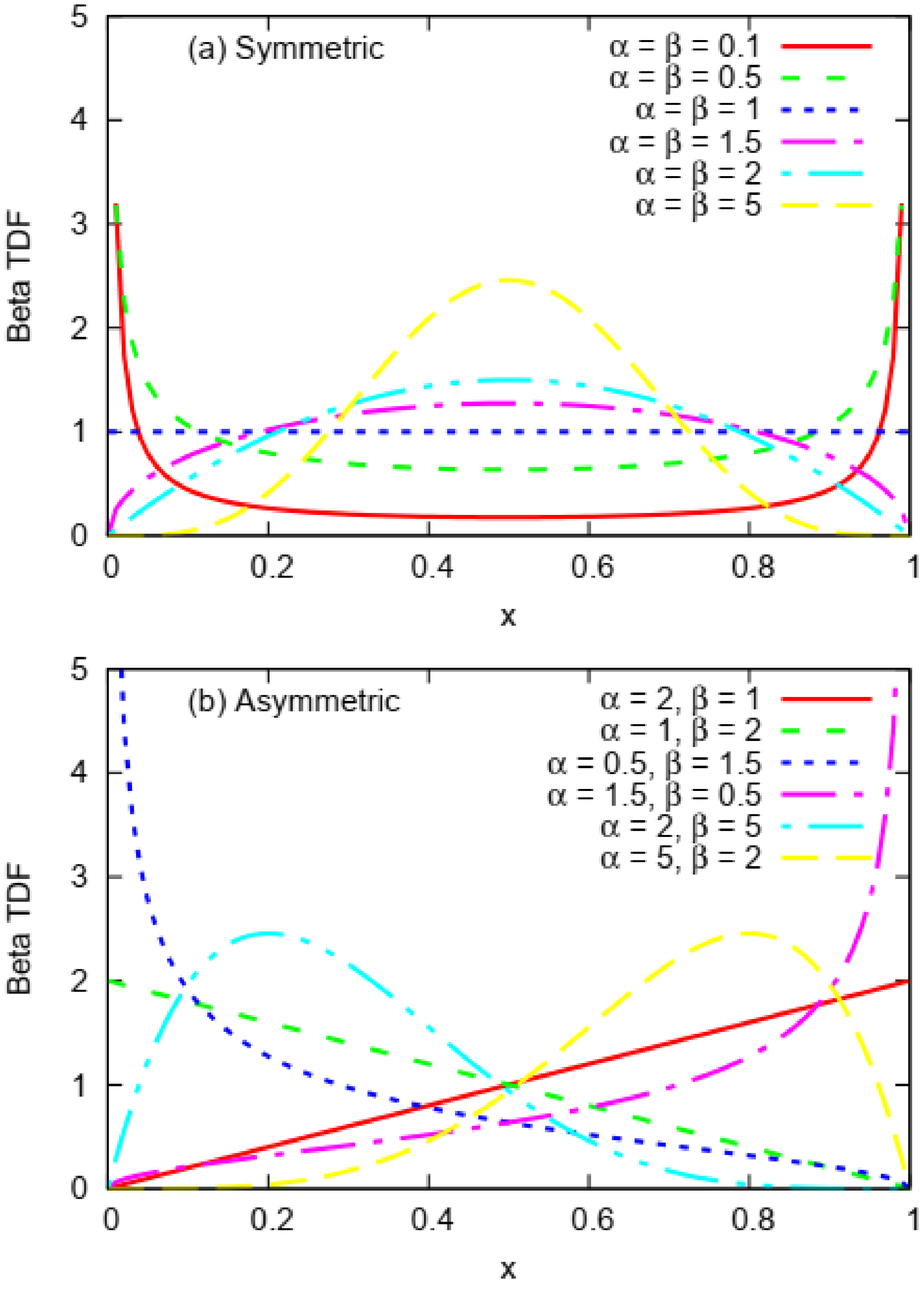

Fig. 2. Symmetric and Asymmetric Beta TDF as a function x with different combination of $\alpha$ and $\beta$.

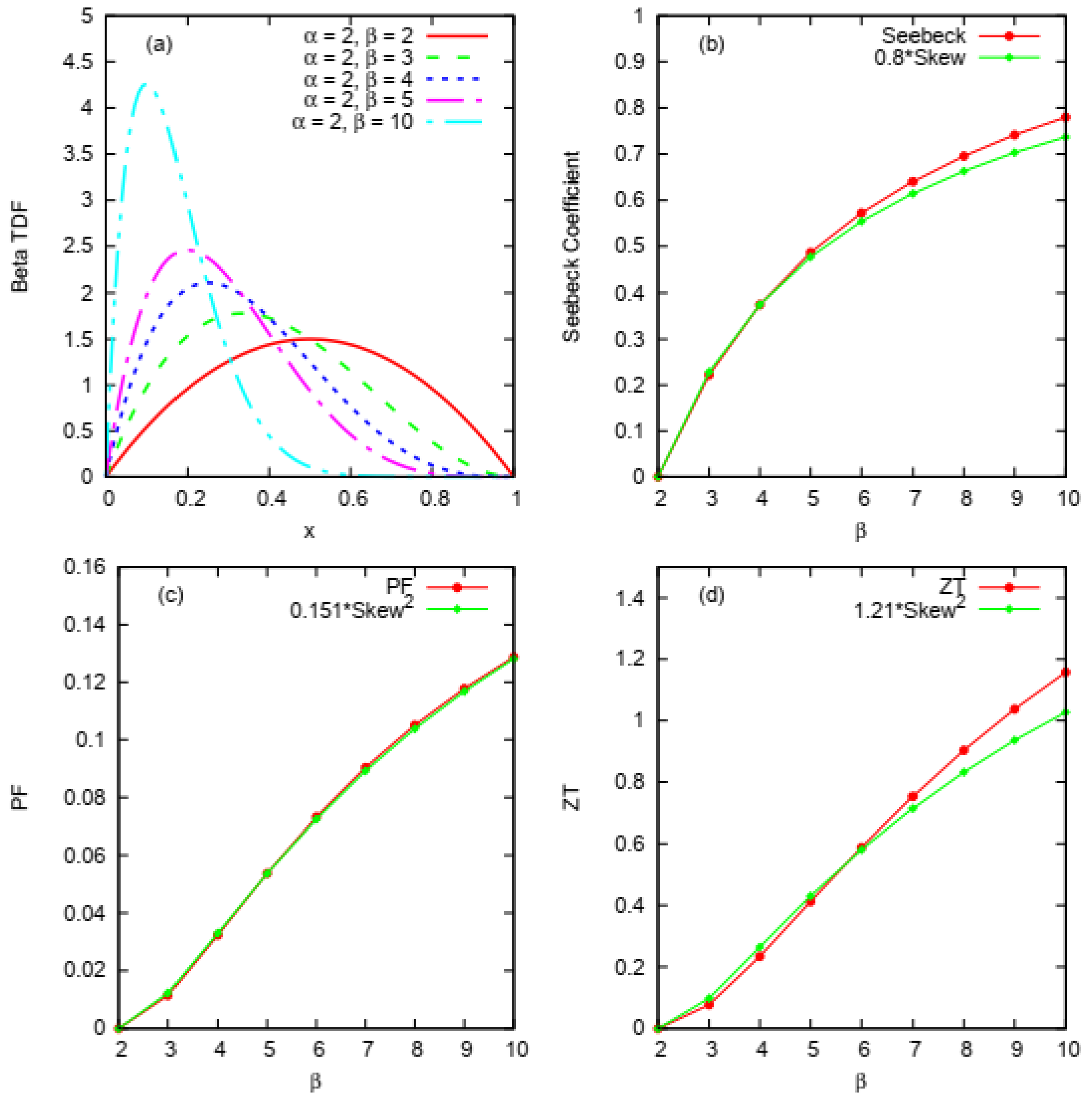

Fig. 3. The comparison of thermoelectric properties and Skewness of Beta TDF as a function of $\beta$. (a) Symmetric and Asymmetric Beta TDF as a function x with fixed α=2 and variable $\beta$. (b) Seebeck coefficient vs 0.8*Skew. (c) PF vs 0.151*Skew$^2$. (d) $ZT$ vs 1.21*Skew$^2$.

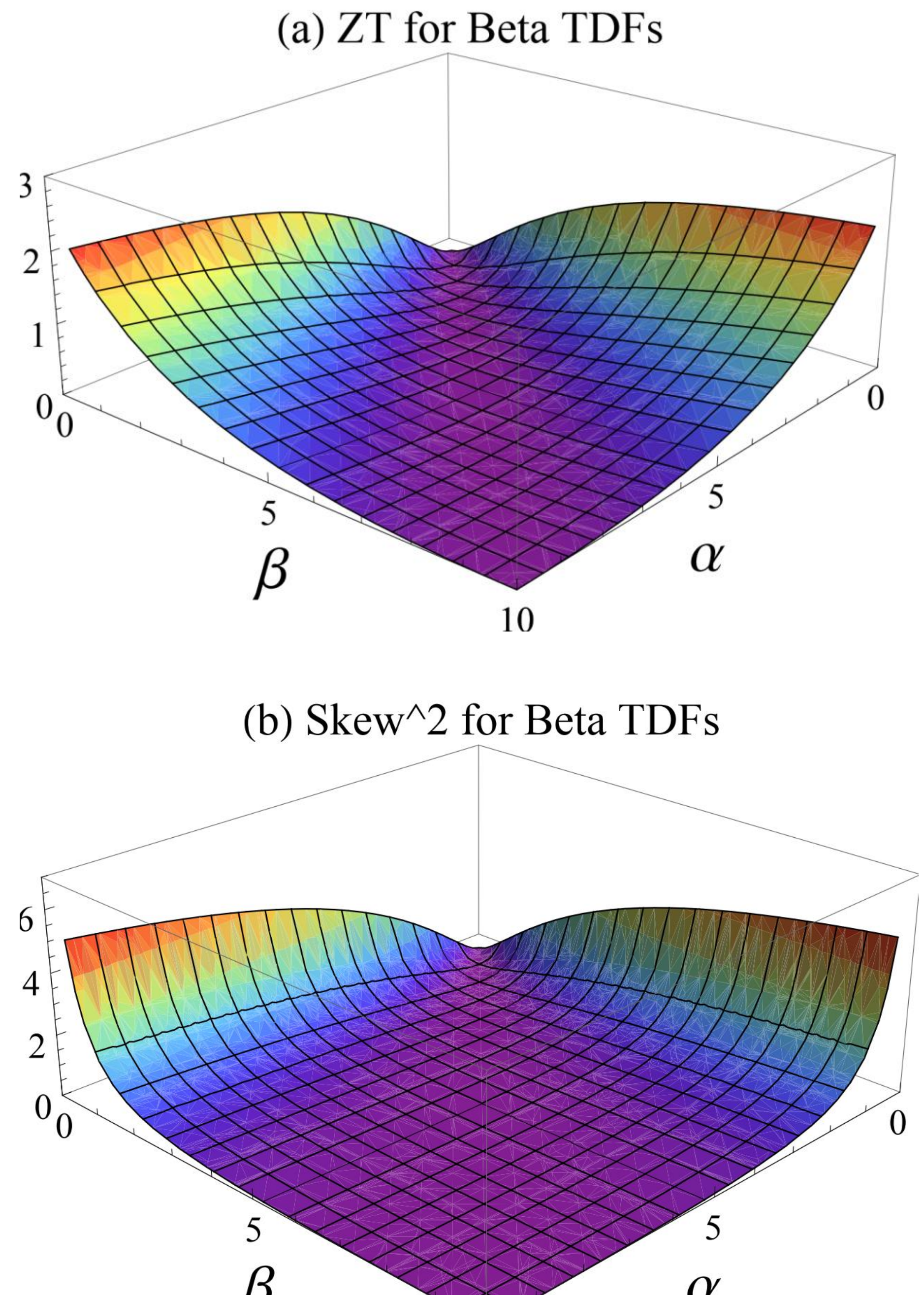

Fig. 4. The comparison of thermoelectric figure of merit (*ZT*) and square of Skewness of Beta TDF in general case for various α and $\beta$. (a) *ZT* (b) Skew$^2$.

## 5. Universal thermoelectric skewness: construction and applications

### 5.1 Construction of universal thermoelectric skewness

Above we already clearly demonstrate that the skewness of TDF, characterization of asymmetry, can be closely related to thermoelectric power (Seebeck coefficient) and other thermoelectric properties. One remaining question is how to apply the concept of skewness in material science, or more specifically, thermoelectric materials. In case a full TDF is available, one may try to fit TDF with Beta function, and then the conventional skewness, as shown in equation (16), can be applied directly to analyze the thermoelectric properties. Unfortunately, in the practice, there are lots of challenges of applying conventional or standard skewness in thermoelectric materials, such as the failure of fitting of TDF by Beta function, or the skewness is undefined for available TDF (for example, for the Cauchy distribution, the mean, skewness, and kurtosis are undefined). At the same time, we also notice that there are several ways to measure skewness and their robustness are often argued. Considering the special characteristics of thermoelectric TDF (e.g., it is meaningful near Fermi energy (or chemical potential at finite temperature, but it will be quickly decayed when the range goes beyond $\pm 5k_BT$, as shown in Fig. 1), it is necessary to construct a novel skewness or asymmetric function for thermoelectric TDF that can be conveniently used by materials science community.

Taking chemical potential as reference, and taking into account both the location parameter and scale parameter, we may construct symmetric and asymmetric functions as,

$$g_S(x, x_0, b) = \frac{1}{2}[g(x, x_0, b) + g(-x, x_0, b)].$$

$$g_A(x, x_0, b) = \frac{1}{2}[g(x, x_0, b) - g(-x, x_0, b)]. \quad (19)$$

The symmetric and asymmetric features of $g_S(x, x_0, b)$ and $g_A(x, x_0, b)$ for $g(x, x_0, b)$ being typical Gaussian are shown in Fig. 5 with $x_0$=1 and $b$=1. Because these functions are symmetric and asymmetric by definition, they are fundamental and universal for distributions with any form of shape.

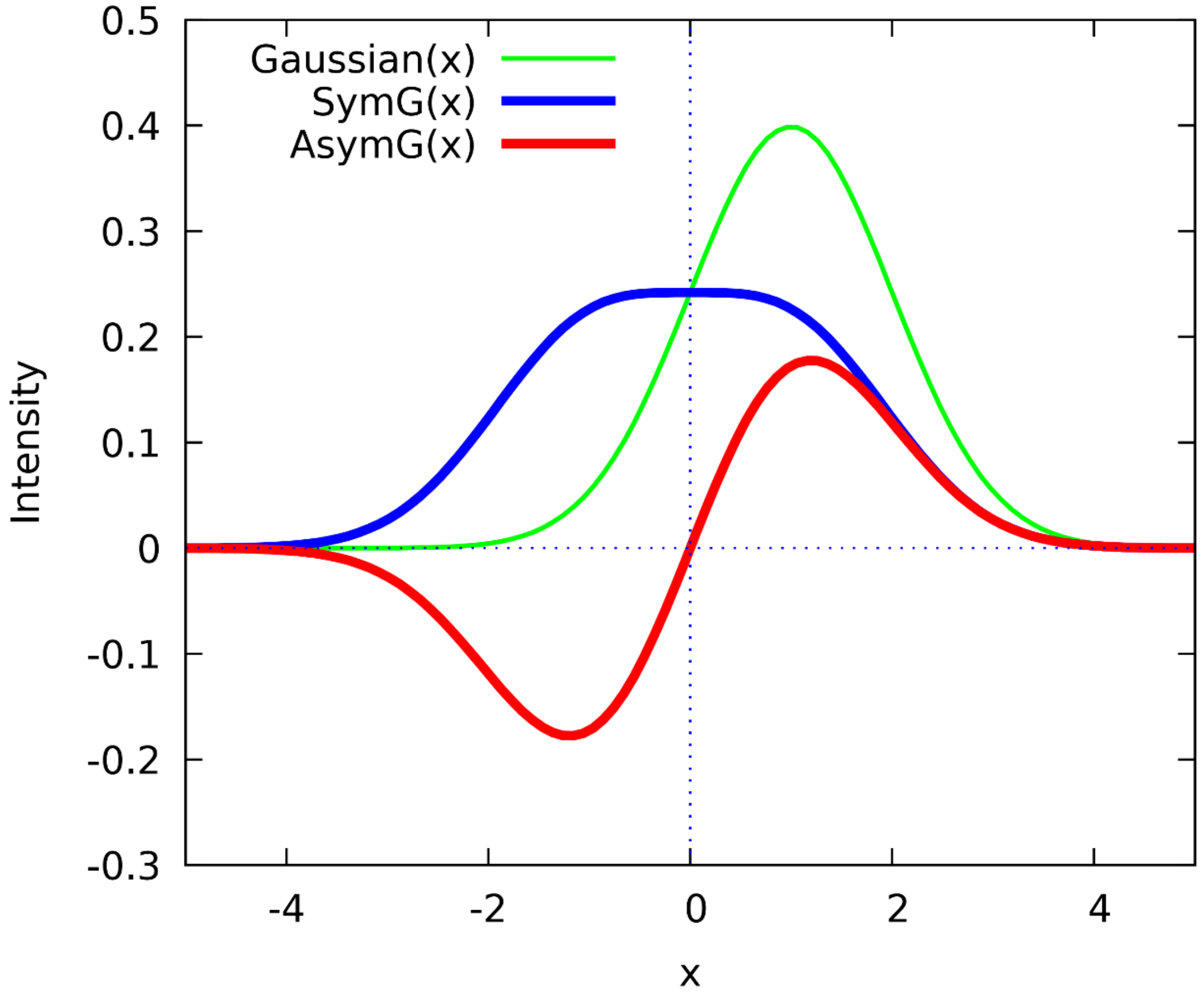

Fig. 5. The symmetric and asymmetric features of $g_S(x,x_0,b)$ and $g_A(x,x_0,b)$. The $g(x,x_0,b)$ in typical Gaussian form ($x_0$=1 and $b$=1) , labeled as "Gaussian(x)", is also shown for comparison. The symmetric function is labeled as "SymG(x)", and the asymmetric function is labeled as "AsymG(x)", respectively.

Then we can define the ***thermoelectric skewness*** as,

$$\gamma^{TE}(x_0,b) = \frac{\int g_A(x,x_0,b)D_1(x,x_0,b)dx}{\int g_S(x,x_0,b)D_0(x,x_0,b)dx} \quad (20)$$

In case only the qualitative trend of thermoelectric skewness (e.g., the indication of improving direction of thermoelectric power) is needed, then integrals can be replaced by finite number of sum (using chosen typical $x_i$),

$$\gamma^{TE}_{Approx,n}(x_0,b) = \frac{\sum_i^n [a_{Ai} g_A(x_{Ai},x_0,b) + c_{Ai}]}{\sum_i^n [a_{Si} g_S(x_{Si},x_0,b) + c_{Si}]} \quad (21)$$

Due to the decay feature of the derivative of the Fermi-Dirac distribution function, the coefficients $a_i$ have larger weight for energy sampling position $x_i$ in the range about $\pm 5k_BT$ around chemical potential, but approach to zero for $x_i$ far away from chemical potential. This interesting feature ensures that the finite number of sum is enough to well reproduce the thermopower but with much simpler form. The simplest case of the ***thermoelectric skewness*** is the ***ratio of asymmetric function and symmetric function*** at typical energy sampling positions, respectively,

$$\gamma^{TE}_{Approx,1}(x_0,b) = \frac{a_{A1} g_A(x_{A1},x_0,b)}{a_{S1} g_S(x_{S1},x_0,b)} \quad (22)$$

where the first approximated parameters $x_{S1}$ and $x_{A1}$ are the values at which $D_0(x)$ and $D_1(x)$ reach their maximum, respectively, namely, $x_{S1}$= 0 and $x_{A1}$= 1.543405. Surprisingly, this simplified thermoelectric skewness, $\gamma^{TE}_{Approx,1}(x_0, b)$, will be shown later to have similar characteristics of third moment of skewness with mean replaced by chemical potential. Therefore, the proposed thermoelectric skewness in this work has many advantages. It is not only having characteristics of conventional skewness, but also convenient to be applied in theoretical and experimental TDF for thermoelectric materials.

### 5.2 The universal thermoelectric skewness vs conventional skewness

Here we will compare the universal thermoelectric skewness and the conventional skewness with third moment as well as the thermoelectric power, to further illustrate the unique characteristics of the universal thermoelectric skewness. At the same time, this is also the showcase for application of universal thermoelectric skewness for prediction of thermoelectric power.

Suppose we have TDF in shifted Gaussian function form with typical scale parameter $b$=2 but adjustable location parameter $x_0$, we can then calculate the thermoelectric power and the skewness. For the Fisher's moment coefficient of skewness (eq. 18), if the skewness is defined about the mean of distribution, then the skewness of Gaussian TDF is zero. However, if we take any value (e.g., the chemical potential) that is away from the mean as the reference point, then the distribution is no longer symmetric about the reference point. The asymmetry and skewness for shifted distribution can therefore be defined accordingly. Take the chemical potential as the reference point, we can then rewrite the conventional skewness with third standardized moment as follows,

$$\gamma_{3,0} = E\left[\left(\frac{X-0}{\mathrm{SD}}\right)^3\right]. \tag{23}$$

Where SD is the standard deviation, and E is the expectation operator. The skewness $\gamma_{3,0}$ is the modified version of conventional skewness.

For shifted Gaussian TDF, the symmetric and asymmetric functions (Eq. 19) can be also written as,

$$g_S(0, x_0, 2) = \frac{1}{2}[g(0, x_0, 2) + g(-0, x_0, 2)],$$

$$g_A(x_{A1}, x_0, 2) = \frac{1}{2}[g(x_{A1}, x_0, 2) - g(-x_{A1}, x_0, 2)]. \tag{24}$$

The thermoelectric skewness can be obtained accordingly,

$$\gamma^{TE}_{Approx,1}(x_0, 2) = \frac{g_A(x_{A1}, x_0, 2)}{g_S(0, x_0, 2)}. \tag{25}$$

It is interesting to see that both conventional skewness $\gamma_{3,0}$ and the thermoelectric skewness $\gamma^{TE}_{Approx,1}(x_0, 2)$ are in good agreement with the thermoelectric power, as shown in Fig. 6. Moreover, the thermoelectric skewness not only perfectly captures the main feature of conventional skewness $\gamma_{3,0}$ (namely, the curvature), but is also be closer to thermoelectric power. This comparison confirms the unique feature of our proposed thermoelectric skewness, and its special role of connection between the statistics of TDF and thermoelectric properties of materials.

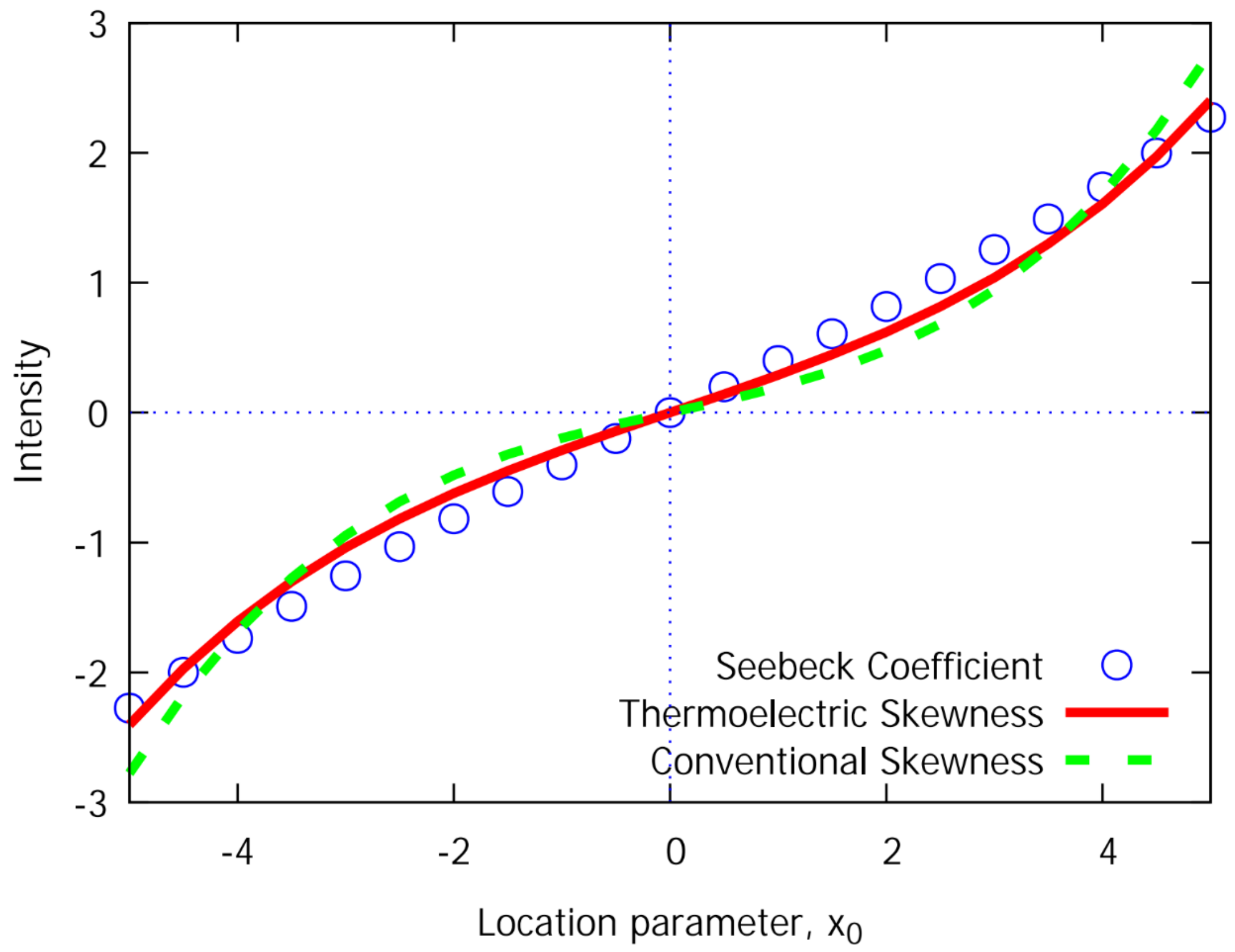

Fig. 6. The comparison of thermoelectric skewness and conventional skewness to Seebeck coefficient (thermoelectric power) using shifted Gaussian TDF. The conventional skewness has been scaled by 0.12.

### 5.3 Application of universal thermoelectric skewness

In this section, we will show how to apply the aforementioned proposed thermoelectric skewness of TDF (namely, asymmetricity of TDF) to interpret the thermoelectric components obtained by either first principles calculations or experimental measurements. The chosen materials cover from amorphous metal alloys (e.g., *a*-$Al_{84}Fe_{16}$), bulk quasicrystals (e.g., *i*-AlCuFe) from experiments and quasicrystal with icosahedral (1/1) approximant from *ab initio* calculations. These examples represent wide range of materials and two typical kinds of characterization of thermoelectric materials, namely, thermoelectric measurements and first principles calculations. Moreover, quasicrystals are also a new class of materials with ordered but not periodic structure, having unique surfaces and properties, as well as wide applications [61-70]. Here, we will use these typical examples to illustrate the modeling of thermoelectric skewness, and make comparison with *ab initio* calculations and experiments.

As shown by Landauro et al [63], the spectral resistivity of icosahedral bulk quasicrystals i-$Al_{62}Cu_{25.5}Fe_{12.5}$, hypothetical i-AlCuFe (1/1) approximant with Cockayne model [62], and amorphous $a$-$Al_{84}Fe_{16}$, can be modeled by Lorentzians. Taking self-consistent Fermi level of the perfect approximant as reference ($\varepsilon_F^{sc} = 0$), and using the same symbols as above cases, the Lorentzians can be expressed as,

$$L(\varepsilon - x_0, b) = (b/\pi)/[(\varepsilon - x_0)^2 + b^2]. \quad (26)$$

Where the location parameter $x_0$ defines the Lorentzian peak position (with reference to $\varepsilon_F^{sc}$), and the height is $1/(\pi b)$. Then the model spectral resistivity can be expressed as combination of two Lorentzians [63] (as shown in Fig. 7a),

$$\hat{\rho}(\varepsilon) = A[L(\varepsilon - x_{01}, b_1) + \alpha L(\varepsilon - x_{02}, b_2)] \quad (27)$$

The location parameters $x_{0i}$ and scaling parameters $b_i$ as well as mixing factor $\alpha$ and unit dependent fitting parameter $A$ are obtained by fitting to either ab initio calculations or experiments. It is found that single Lorentzian is sufficient for amorphous $a$-$Al_{84}Fe_{16}$ (see insert figure of Fig.7a), while for icosahedral bulk quasicrystals i-$Al_{62}Cu_{25.5}Fe_{12.5}$, and hypothetical i-AlCuFe (1/1) approximant, two Lorentzians are necessary (spectra are shown in Fig.7a and Fig.7b, and detailed parameters [63] are presented in Table S1).

The model spectral resistivity (Fig.7b) is related to spectral conductivity through

$$\hat{\rho}(\varepsilon) = \hat{\sigma}^{-1}(\varepsilon). \quad (28)$$

In order to directly compare our thermoelectric skewness and related analysis with ab initio results and experimental measurement, in this section, we will directly utilize thermoelectric transport coefficients from Eq. (2-5) for this purpose. Caution should be made when utilizing thermoelectric skewness (Eqs. 21, 22), and the parameters $x_{A1}$ and $x_{S1}$ should be scaled by $k_BT$, in case the dimensionless quantities are no longer used in thermoelectric transport coefficients. Therefore, the parameters $x_{A1}(T)$ and $x_{S1}(T)$are becoming temperature dependent.

The thermoelectric powers of quasicrystals are then calculated by both Boltzmann transport equation (Boltz, Eq. (2-5)) and that from thermoelectric skewness (Eq. 22) as a function of chemical potential and compared in Fig. 7c. It is clearly seen that the thermoelectric skewness is able to capture the trend of thermoelectric power perfectly with respect to the variance of chemical potential. Similar excellent agreements are also found for the cases of amorphous $a$-$Al_{84}Fe_{16}$ and hypothetical i-AlCuFe (1/1) approximant. Especially, the locations of chemical potential of bulk quasicrystal that leads to zero thermoelectric power or maximum value, as well as the sign-changed behavior of thermoelectric power predicted by BTE can be well reproduced by our thermoelectric skewness.

Finally, we compare our calculated temperature-dependent thermoelectric power by BTE and thermoelectric skewness with experiments [64], as shown in Fig. 7d. In this calculation, the temperature dependent chemical potential from ref. [63] is adapted,

$$\mu(T) \approx \varepsilon_F - \xi T^2. \quad (29)$$

Where $\varepsilon_F$ is Fermi energy at 0 K for i-$Al_{62}Cu_{25.5}Fe_{12.5}$ (labeled as "A") and i-$Al_{63}Cu_{25}Fe_{12}$ (labeled as "B"). Using the parameters $\varepsilon_F(A)$ = 0.223 eV, $\varepsilon_F(B)$ = 0.242 eV, and $\xi$ = 0.10×10$^{-6}$ eV/K$^2$, the agreement between theory and experiment is reasonably well. It further confirms that the proposed thermoelectric skewness is not just an alternative definition of skewness for statistics, but more importantly, it can well capture the key features of thermoelectric power, although only simplified analytical expression is used. It is worth to note, that although amorphous metal alloys and quasicrystals are used for examples in this work, in principle, our proposed thermoelectric skewness of TDF can be applied for any thermoelectric materials if full or part of TDF are available. Therefore, we consider our thermoelectric skewness is universal.

Our results are also very useful in practical searching for better thermoelectric materials. For example, it has been shown that the band engineering through the use of the thallium impurity levels in lead telluride (PbTe) enhances the thermoelectric performance [71]. Based on our findings in this work, the enhancement of thermoelectric performance found in [71] can be well understood by comparing DOS above Fermi energy and that below Fermi energy. PbTe doped with Tl introduce an additional DOS peak above the Fermi energy, by using our asymmetric function or thermoelectric skewness, with TDF being replaced by experimental DOS, the asymmetry or skewness of DOS is enhanced, and thus the thermoelectric performance will be increased accordingly.

In general, our work suggests that, to search for better thermoelectric materials, one should first check the asymmetric feature of TDF. If full TDF is not available, it will also be useful to check the asymmetry of the components of TDF, such as DOS, relaxation time, and carrier velocity for the first screening, if they are available, either by measurements or computations.

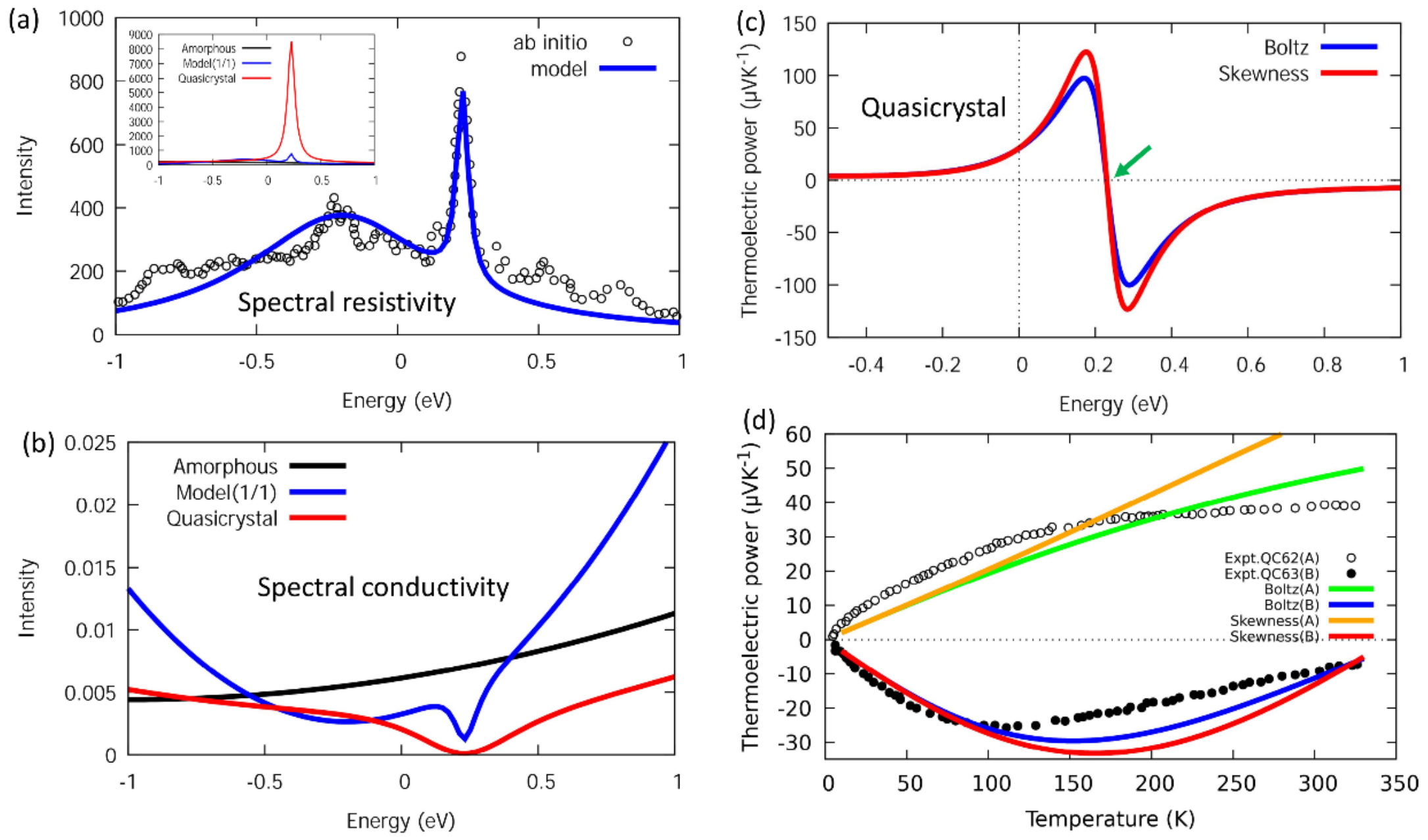

Fig. 7. The spectral resistivity, spectral conductivity (namely, TDF), and thermoelectric power of quasicrystals with different phases. (a) The spectral resistivity of hypothetical i-AlCuFe (1/1)

approximant obtained from ab initio calculations (open circle) and modelling with two Lorentzians (solid line) [63]. (b) The spectral conductivity of amorphous $a$-$Al_{84}Fe_{16}$, while for icosahedral bulk quasicrystals i-$Al_{62}Cu_{25.5}Fe_{12.5}$, and hypothetical i-AlCuFe (1/1) approximant. (c) The thermoelectric power of bulk quasicrystals i-$Al_{62}Cu_{25.5}Fe_{12.5}$ obtained from Boltzmann transport equation (Boltz) and that from thermoelectric skewness (proposed in this work) as a function of chemical potential at 300 K. (d) The thermoelectric power of bulk quasicrystals as a function of temperature. The experimental thermoelectric power [64] of i-$Al_{62}Cu_{25.5}Fe_{12.5}$ (labeled as "A") and i-$Al_{63}Cu_{25}Fe_{12}$ (labeled as "B") are presented with open circles and filled circles, respectively. The theoretical thermoelectric power of bulk quasicrystals obtained from Boltzmann transport equation (Boltz) are shown with green and blue lines for A and B, respectively. The thermoelectric skewness of A and B are illustrated with orange and red lines and scaled by $a_{A1}/a_{S1}$=183.

## 6. Conclusion

In conclusion, we have systematically investigated the relationship between the parameters of TDFs (including location parameter, scale parameter, and shape parameter) and thermoelectric properties. We found that the asymmetry of TDF, characterized by skewness, is a more general and universal parameter to capture the trend of thermoelectric power, power factor as well as *ZT*. By taking TDF being Beta distribution as a showcase, it is found that the Seebeck coefficient varies closely with the first order of skewness, while the power factor and *ZT* are well fitted with the second order of skewness (namely, the square of skewness). The finding of close relationship between shape of TDF and thermoelectric performance is striking and surprising. Firstly, for any form of TDF, if it is symmetric about its mean, then the thermoelectric power (Seebeck coefficient), power factor, and *ZT* are zero for the case of the chemical potential $\mu/(k_BT)$ being set to the mean of the distribution. Secondly, the asymmetry of TDF, characterized by skewness, is a universal parameter to describe the trend of thermoelectric properties for the shape of TDF. Thirdly, the thermoelectric transport properties including thermoelectric figure of merit, which are based on typical quantum statistics (Fermi-Dirac distributions), can be well described by typical shape parameter (skewness) for classical statistics.

In many cases, the skewness is undefined (for example, the Cauchy distribution or Lorentzian distribution) or difficult for use in materials science. It is necessary to construct a new skewness that can be adapted for materials science. We then proposed a novel skewness, so-called ***thermoelectric skewness***, constructed from symmetric and asymmetric TDF functions. It is demonstrated that the proposed thermoelectric skewness is not only to perfectly capture the main feature of conventional skewness, but also to predict accurately the thermoelectric power, by comparing with ab initio calculations and experiments. This comparison confirms the unique feature of our proposed thermoelectric skewness, and its special role of connection between the statistics of TDF and thermoelectric properties of materials.

Our results indicate that it is possible to choose suitable parameters from classical statistics to capture some important features of quantum statistics, thus the quantum statistics and classical statistics can be linked or bridged by some typical parameters that characterizing the shape of distribution. This may suggest that large amount of information or tools in classical statistics or probability theory may be applied for quantum statistics and used for optimization of important properties (such as the thermoelectric power, the power factor, and *ZT*) of quantum systems. Our results are also very useful in practical searching for better thermoelectric materials.

**Data Availability**

The data that support the findings of this study are available from the corresponding author upon reasonable request.

**Conflicts of Interest**

There are no conflicts to declare.

**Acknowledgements**

This research was supported by Xiamen University Malaysia Research Fund (No. XMUMRF/2019-C3/IORI/0001).

**Supplementary Materials**

S1. Thermoelectric components for the constant TDF.

S2. TDF with constant location-scale family: the case of Dirac delta function.

S3. TDF with typical general location-scale family: TDF in Gaussian function form.

S4. Parameters of Lorentzians for spectral resistivities. (*Supplementary Materials*)

Supplemental Materials for:

# Asymmetrical transport distribution function: skewness as a key to enhance thermoelectric performance

Jin-Cheng Zheng[(a,b)]*
(a) Department of Physics, Xiamen University, Xiamen 361005, China.
(b) Department of Physics and Department of New Energy Science and Engineering, Xiamen University Malaysia, Sepang 43900, Malaysia.
* Electronic mail: jczheng@xmu.edu.cn

## S1. Thermoelectric components for the constant TDF.

For the case of thermoelectrics with constant TDF, namely, $g(x) = g_0$, the dimensionless integrals $J_n$ becomes

$$J_n = g_0 \int_{-\infty}^{+\infty} D_n(x)dx. \tag{s1}$$

Then we have, $J_0 = g_0$, $J_1 = 0$, and $J_2 =$ $J_2 = \frac{\pi^2}{3} g_0$. The reduced electrical conductivity is $\sigma_r = g_0$. The reduced electronic thermal conductivity is $\kappa_{e,r} = \frac{\pi^2}{3} g_0$. The Seebeck coefficient $S_r$, power factor and the $ZT$ are all zero.

## S2. TDF with constant location-scale family: the case of Dirac delta function.

The TDF in single Dirac delta function form can be expressed as,

$$g(x) = g_0 \delta(x - x_0), \tag{s2}$$

where $g_0$ is constant. This TDF is actually a Delta shape peak located at the $x =$ $x_0$ position away from Fermi energy level. Taking advantage of *sifting property* of the delta function (the integral of the time-delayed Dirac delta),

$$\int_{-\infty}^{+\infty} f(x)\delta(x - x_0)dx = f(x_0), \tag{s3}$$

one can obtain integrals $J_n$ and transport components as,

$$J_n = g_0 D_n(x_0). \tag{s4}$$

Thus, the reduced electric conductivity is,

$$\sigma_r = g_0 D_0(x_0), \tag{s5}$$

the reduced thermoelectric power (Seebeck Coefficient) becomes

$$S_r = x_0, \tag{s6}$$

and the reduced power factor can be obtained as

$$P_r = g_0 x_0^2 D_0(x_0). \tag{s7}$$

Because the reduced electronic thermal conductivity is 0,

$$\kappa_{e,r} = 0, \tag{s8}$$

the figure of merit (*ZT*) is finally obtained as

$$ZT = g_0 x_0^2 D_0(x_0)/\gamma. \tag{s9}$$

The optimal value of thermoelectric components requires different optimal location parameter $x_0$. When the location parameter $x_0 = 0$, the reduced electric conductivity has the highest value of $\sigma_r = g_0/4$, because the $D_0(x_0=0)$ is the maximum. However, for the reduced power factor and *ZT*, the optimal value of location parameter $x_0$ is shifted from 0. The function of $x_0^2 D_0(x_0)$ reaches maximum 0.44 at $x_0 \approx 2.40$, and the maximum of reduced power factor of 0.44 $g_0$ and *ZT* of 0.44 $g_0/\lambda$ can be obtained accordingly. After the optimization of location parameter $x_0$, the *ZT* can be further enhanced by the increase of scale parameter $g_0$, or by the reduction of lattice thermal conductivity. By assuming a typical lattice thermal conductivity, $\kappa_l = 1$ $Wm^{-1}K^{-1}$ at room temperature, we have $\lambda = 0.0976$, then the optimal *ZT* can be as high as 4.50 $g_0$ for general Delta function as TDF. For the simpler case of TDF being the normalized Delta function ($g_0 = 1$), the *ZT* is 4.50 at room temperature, which is more than 4 times than the *ZT* of the best thermoelectric material, $Bi_2Te_3$ at room temperature. For the cases when the scale parameter $g_0 > 1$, and lattice thermal conductivity is lower than 1 $Wm^{-1}K^{-1}$, the maximum *ZT* will be even much higher (e.g., if $g_0 > 2.23$, *ZT* is higher than 10).

**S3. TDF with typical general location-scale family: TDF in Gaussian function form**

The Gaussian TDF thus can be expressed as,

$$g(x) = \frac{1}{b\sqrt{2\pi}} \exp\left[-\frac{(x-x_0)^2}{2b^2}\right], \tag{s10}$$

The location parameter $x_0$ defines the peak position of the Gaussian function, and the scale parameter *b* characterizes the width of the peak. The TDFs in Gaussian function form with typical location parameter ($x_0 = 0$ and 2.4) and scale parameter (b = 0.5, 1, and 3) are shown in Fig. S1(a). The thermoelectric properties, including the reduced electrical conductivity, the reduced thermoelectric power (Seebeck coefficient), the power factor, the reduced electron thermal conductivity and the figure of merit (*ZT*) of thermoelectric material as a function of location parameters and scale parameters are shown in Fig. S1(b-f), respectively. Typical location parameters ($x_0 = 0$, 1, 2.4 and 5) are shown to illustrate the trend of thermoelectric properties.

The Gaussian TDF represents a wide range of general distributions. Firstly, regarding the scale parameter ($b$), it represents the width of the peak or the spread of the distribution. The smaller the value of $b$, the narrower or shaper the peak; the larger the value of $b$, the flatter the peak. This feature is similar with density of states (DOS) of thermoelectric materials. For more localized orbital, the peak of DOS is narrower, and for more delocalized electrons, the associated DOS is more flat. In extreme cases, such as if the scale parameter ($b$) approaches zero, the TDF will be approaching Dirac delta function, $\delta(x)$. On the other hand, if the scale parameter ($b$) approaches infinite, then the TDF will be approaching constant distribution. Both extreme cases have been discussed in previous parts. Secondly, the location parameter can be associated with the position of Fermi energy at zero temperature or chemical potential at finite temperature. It is well known that the chemical potential can be tuned by doping [S1] or pressure [S2,S3] or strains [S4,S5], thus the carrier velocity, relaxation time and DOS can be modified accordingly. Subsequently, the thermoelectric transport properties are changed too. Therefore, if the scale parameter ($b$) reflects the peak width of TDF (mainly related to electronic structure features such as DOS and its spread associated with temperature through Fermi-Dirac distribution), the location parameter characterizes the shifting the chemical potential due to chemical pressure (such as impurity or defects) or mechanical pressure (including hydrostatic pressure or strain fields).

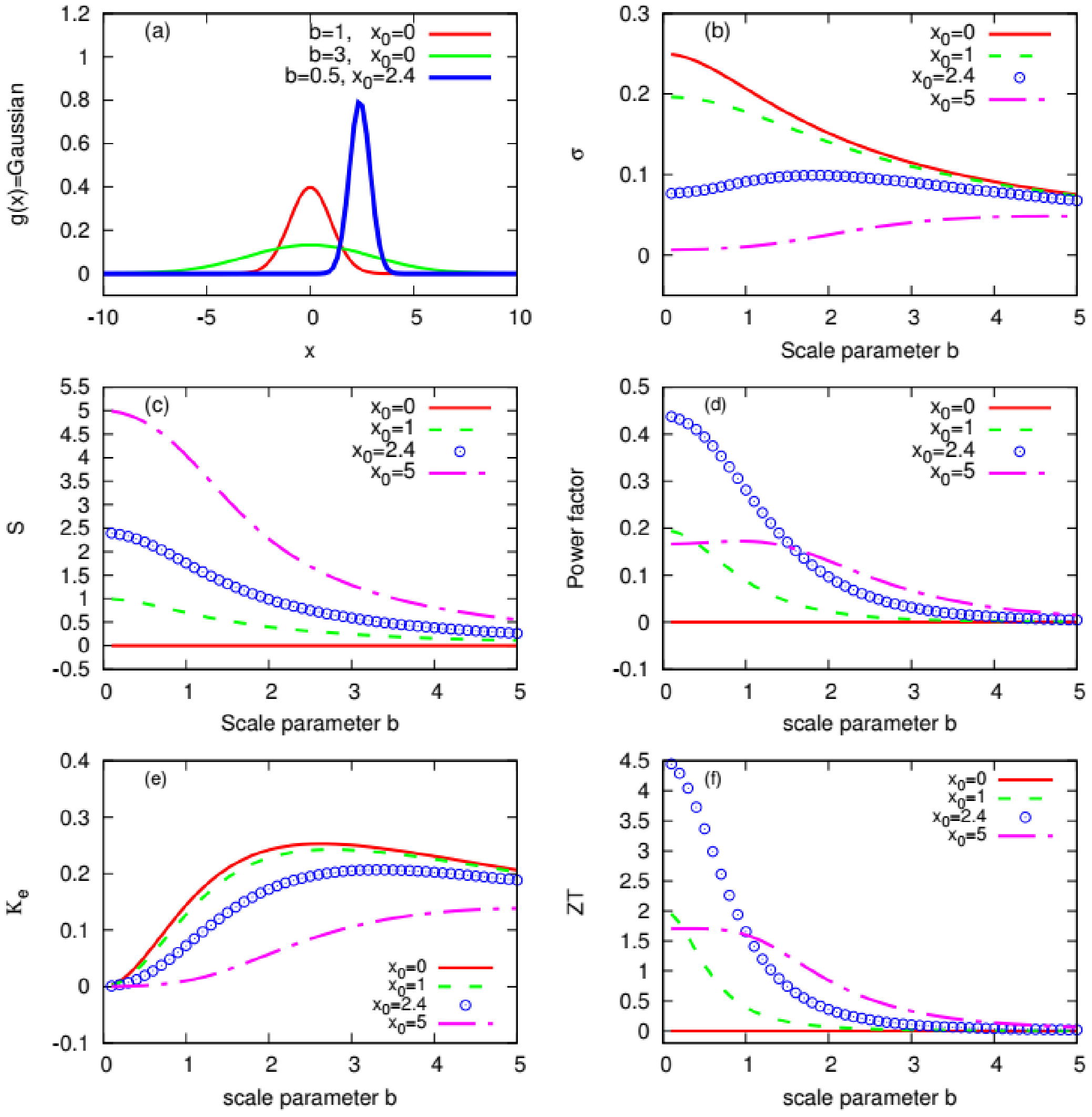

Fig. S1. The thermoelectric properties of system with TDF in a Gaussian function form. (a) The transport density function (TDF) in a Gaussian function form. Examples of the location parameter ($x_0$=0 and 2.4) and scale parameter ($b$= 0.5, 1, and 3) are shown in the figure. The reduced electrical conductivity (b), the reduced thermoelectric power (Seebeck coefficient) (c), the power factor (d), the reduced electron thermal conductivity (e) and the figure of merit (*ZT*) (f) of thermoelectric material as a function of location parameters and scale parameters. Typical location parameters ($x_0$=0, 1, 2.4 and 5) are presented.

From Fig. S1, it can be found that the response of thermoelectric components to location parameter and scale parameter are significantly different. The electrical conductivity and the electron thermal conductivity decrease with the increase of location parameter $x_0$, but the thermoelectric power increases with increase of location parameter. When the location parameter $x_0$ = 0, the mean, median and mode of the Gaussian distribution are located at 0. The electrical conductivity and the electron thermal conductivity reaches its largest value, while for the thermoelectric power (Seebeck coefficient, S), it is just the opposite, and it becomes zero, independent on scale parameter. Therefore, other thermoelectric quantities, which are proportional to thermoelectric power (S), such as the power factor, and *ZT*, are all zero for the case of the location parameter $x_0 = 0$, as shown in solid line in red color in Fig. S1 (c, d, and f). This is very similar with (or approximately close to) the case of good metal in real materials. For a good metal, the DOSs located at the Fermi energy (or chemical potential at finite temperature) are normally quite large, which can be very close to Gaussian TDF with location parameter near zero. The trends of the electrical conductivity and the electron thermal conductivity as well as the thermoelectric power in terms of location parameters well explain that the normal good metals have very good electrical conductivity and electron thermal conductivity, but have very poor thermoelectric power.

For semiconducting materials, the DOS may have several peaks, and the peak positions are normally away from the Fermi energy depending on doping level. This case is similar with the Gaussian TDF with non-zero location parameter and finite scale parameter. From Fig. S1, one can see that with the increase of location parameter, the electrical conductivity reduces but the thermoelectric power increase. It makes the trend of power factor and *ZT* varies more complicated with location parameter and scale parameter, because the power factor and *ZT* are strongly coupled with both electrical conductivity and thermoelectric power. The overall absolute value of power factor or *ZT* at larger scale parameter are very small although they will be slightly enhanced by increasing location parameter. This is because Gaussian TDF with large scale parameter will be closer to constant TDF, and as we discussed in previous part, the electrical conductivity can be maintained in certain level, but the thermoelectric power, power factor and thus the *ZT* are becoming small. To enhance the *ZT*, the scale parameter should be reduced and the value of location parameter should be optimized. As shown in Fig. S1, the optimal location parameter is about $x_0 = 2.4$, the power factor and *ZT* are significantly enhanced with smaller scale parameter, and they reach the maximum when the scale parameter approaches zero. That is exactly the case of TDF being Dirac delta function.

**S4. Parameters of Lorentzians for spectral resistivities.**

Table S1. Parameters of Lorentzians for spectral resistivities, values given by Landauro et al [S6].

| Materials | A | $x_{01}$ | $b_1$ | $\alpha$ | $x_{02}$ | $b_2$ |
|---|---|---|---|---|---|---|
| | (μΩ cm eV) | (eV) | (eV) | | (eV) | (eV) |
| amorphous *a*-$Al_{84}Fe_{16}$ | 1138.83 | -1.0 | 1.6 | - | - | - |
| quasicrystals i-$Al_{62}Cu_{25.5}Fe_{12.5}$ | 1047 | -0.2 | 1.35 | 1.0 | 0.23 | 0.04 |
| i-AlCuFe (1/1) approximant | 477.25 | -0.2 | 0.4 | 0.1 | 0.23 | 0.025 |